\documentclass[preprint,astrosymb]{aastex631}

\shorttitle{Extragalactic Star Clusters with Roman and Rubin}
\shortauthors{Extragalactic Star Clusters with Roman and Rubin}

\begin{document}
\title{Extragalactic Star Cluster Science with the Nancy Grace Roman Space Telescope's High Latitude Wide Area Survey and the Vera C. Rubin Observatory}
\author{Authors: Kristen C. Dage (McGill University, Canada, kcdage@msu.edu)}
\author{Christopher Usher, (The Oskar Klein Centre/Stockholm University, Sweden)}
\author{Jennifer Sobeck, (CFHT, USA)}
\author{Ana L. Chies-Santos (UFRGS, Porto Alegre, Brazil)} 
\author{R\'obert Szab\'o (Konkoly Observatory, Hungary)}
\author{Marta Reina-Campos (McMaster University, Canada)}
\author{Léo Girardi (INAF-Padova, Italy)}
\author{Vincenzo Ripepi (INAF-Capodimonte, Italy)}
\author{Marcella Di Criscienzo (INAF-Rome, Italy)} 
\author{Ata Sarajedini (Florida Atlantic University, USA)}
\author{Will Clarkson (University of Michigan-Dearborn, USA)}
\author{Peregrine McGehee (SLAC, USA)}
\author{John Gizis (University of Delaware, USA)}
\author{Katherine Rhode (Indiana University, USA)}
\author{John Blakeslee (NOIRLab, USA)}
\author{Michele Cantiello (INAF-Abruzzo, Teramo, Italy)}
\author{Christopher A. Theissen (University of California, San Diego, USA)}
\author{Annalisa Calamida (Space Telescope Science Institute, Baltimore, USA)}
\author{Ana Ennis (University of Waterloo, Canada)}
\author{Nushkia Chamba (The Oskar Klein Centre/Stockholm University, Sweden)}
\author{Roman Gerasimov (University of California, San Diego, USA)}
\author{R. Michael Rich (University of California, Los Angeles, USA)}
\author{Pauline Barmby (Western University, Canada)}
\author{Annette M. N. Ferguson (University of Edinburgh, UK)}
\author{Benjamin F. Williams (University of Washington, USA)}
\begin{abstract}

 The Nancy Grace Roman Telescope's High Latitude Wide Area Survey will have a number of synergies with the Vera Rubin Observatory's Legacy Survey of Space and Time (LSST), particularly for extragalactic star clusters. Understanding the nature of star clusters and star cluster systems are key topics in many areas of astronomy, chief among them stellar evolution, high energy astrophysics, galaxy assembly/dark matter, the extragalactic distance scale, and cosmology. One of the challenges will be disentangling the age/metallicity degeneracy because young ($\sim$Myr) metal-rich clusters have similar SEDs to old ($\sim$Gyr) metal-poor clusters. Rubin will provide homogeneous, $ugrizy$ photometric coverage, and measurements in the red Roman filters will help break the age-metallicity and age-extinction degeneracies, providing the first globular cluster samples that cover wide areas while essentially free of contamination from Milky Way stars. Roman's excellent spatial resolution will also allow measurements of cluster sizes.\textbf{ We advocate for observations of a large sample of galaxies with a range of properties and morphologies in the Rubin/LSST footprint matching the depth of the LSST Wide-Fast-Deep field $i$ band limit (26.3 mag), and recommend adding the F213 filter to the survey. }
\end{abstract}
\section*{scientific categories}
stellar physics and stellar types; stellar populations and the interstellar medium; galaxies. additional keywords: multiwavelength, X-rays.

\section{Introduction}
\label{section:intro}
\vspace{-1mm}

Star clusters provide a window onto a wide range of astrophysical processes and hold the answers to key questions regarding how stars form, evolve, and die, as well as how galaxies form and evolve. Moreover, observational and theoretical studies have demonstrated that globular clusters are effective probes of the shape and structure of their host galaxies, including the galaxies’ dark matter halos \citep{2006ARA&A..44..193B, 2020rfma.book..245B, 2022MNRAS.513.3925R, 2023MNRAS.521.6368R}. Synergistic observations from the Nancy Grace Roman Space Telescope’s High Latitude Wide Area Survey (HLWA) and the Vera C. Rubin Observatory’s Legacy Survey of Space and Time (LSST) will allow for an entirely new characterization of extragalactic star clusters, reaching the turnover of the cluster luminosity function at distances out to 100 Mpc. 


By combining data from the two surveys, we can analyze stellar populations across a wide wavelength range, from Rubin/LSST's $u$-band at 0.38 $\mu$m, to Roman's F184 at 1.84 $\mu$m (or possibly F213 at 2.13 $\mu$m). Moreover, the high spatial resolution of Roman offers the opportunity to estimate the sizes of globular clusters in external galaxies. 
The measurements obtained through Rubin's blue optical filters play a crucial role for globular cluster science.  Relying solely on near-infrared photometry is insufficient to overcome the age-metallicity and age-extinction degeneracies, which pose fundamental challenges in studying unresolved stellar populations. 

While observing extragalactic globular clusters using Roman alone allows us to estimate their metallicities, effectively resolving the age/metallicity degeneracy necessitates data from both Rubin and Roman. Rubin's comprehensive, homogeneous $ugrizy$ photometric coverage, when combined with Roman's measurements in red/near-IR filters and Rubin's measurements in blue optical filters, will extend the temperature, age and colour sensitivity, effectively breaking the age-metallicity and age-extinction degeneracies.

\vspace{-1mm}

\section{Science Areas Benefiting from Roman Observations}
\vspace{-1mm}

The core synergies between the two facilities will drastically reduce the contamination from Milky Way stars and background galaxies present in globular cluster catalogs \citep{2014ApJS..210....4M}, thereby allowing more accurate studies of stellar populations and host galaxy assembly. They will also help us understand the populations of exotic stellar objects found within globular clusters (and by extension, the progenitor sources of gravitational wave events). In the planned 1400 square degree survey, both telescopes expect to detect  $\sim4$ million GCs in F106, F129 and F158 plus about 1.5 million in F184 (assuming AB depths of 26.7 mag in F106, F129 and F158 and 26.0 mag in F184). 

\vspace{-1mm}

\subsection{Galaxy Assembly and Dark Matter}

\vspace{-1mm}

Breaking the age/metallicity degeneracy will enable more effective selection and study of extragalactic star clusters and will represent a major step forward in our understanding of the assembly history of galaxies and their dark matter content.
By inverting the galaxy mass-metallicity relationship, we can use the age and metallicity of a globular cluster to estimate the mass of the host galaxy when the globular cluster formed.
Through such estimates for a number of globular clusters, the assembly history of a galaxy can be studied and compared to the predictions of galaxy formation models \citep{2019MNRAS.486.3134K}. Globular clusters also correlate spatially with tidal streams in stellar halos \citep[e.g.][]{2019MNRAS.484.1756M} hence they can be used to uncover recent accretion events, even when the associated stellar debris is too faint to be detected.  
The large, volume-limited sample of galaxies provided by combining observations from Rubin and Roman will allow a powerful model comparison.
Additionally, the properties of observed globular cluster systems can be compared to the predictions of globular cluster formation and evolution models \citep[e.g.][]{2014ApJ...796...10L, 2018MNRAS.475.4309P, 2021MNRAS.505.5815V, 2023MNRAS.521..124R}.

The total mass of a globular cluster system is very tightly linked to the host galaxy's dark matter halo mass \citep[e.g.][]{2015ApJ...806...36H}, and the spatial distribution of the cluster system  constrains  the shape, profile and orientation of the dark matter halo \citep{2018MNRAS.477.3869H,2022MNRAS.513.3925R,2022arXiv220411861R}. Given the ubiquity of globular clusters in intermediate/high-mass galaxies, they provide an excellent approach to improve our understanding of galaxy assembly and dark matter halos. These links are poorly constrained for low-mass galaxies due to the low number of globular clusters per galaxy and the difficulty of measuring their halo masses. {\it The Rubin-Roman data-set would test the mass regime in which the relationship between cluster properties and dark matter might break down, thus signaling the presence of new physics and addressing open questions such as the minimum halo mass in which star formation can occur.}  Both Rubin and Roman Observatories anticipate excellent control of scattered light, and thus the globular cluster populations can be compared with extended low surface brightness structures associated with galaxies and galaxy clusters.



\subsection{Extragalactic distance scale}

Old globular clusters have been shown to be effective for measuring extragalactic distances with a statistical uncertainty of approximately 10\% for individual galaxies. To achieve this, it is necessary to observe a minimum number of $\sim$100 globular clusters and sample the luminosity function to a depth of at least 1 magnitude fainter than the turnover magnitude \citep[e.g.][]{2010ApJ...717..603V}. By combining the resolution, depth, and near-infrared photometry of Roman with the optical data from the Rubin Observatory, it will potentially be possible to obtain such distances for elliptical galaxies as well as (edge-on) spiral galaxies. This approach will serve as a valuable alternative to the traditional Cepheids+SNeIa method for determining $H_0$ independently or to validate and calibrate distances obtained from other indicators (such as surface brightness fluctuations, SBF and the tip of the red giant branch method) applied to the same galaxies.
\vspace{-1mm}

\subsection{Stellar Exotica/High Energy Astrophysics}

\vspace{-1mm}

Understanding stellar exotica in extragalactic globular clusters has been hampered by the lack of a homogenous data-set which includes ages, metallicities and sizes. Roman’s exquisite spatial resolution will allow us to extract information about the sizes and structures of the globular clusters; this information, when paired with age/metallicity measurements, can be combined with multiwavelength observations to securely link globular cluster properties to stellar exotica, such as low-mass X-ray binaries, fast radio bursts, and ultraluminous X-ray sources  (e.g. \citealt{Dage2021,2022Natur.602..585K}).  Globular clusters are thought to be one of the primary formation sites for the merging black hole-black hole binaries that have been observed by LIGO/Virgo/KAGRA \citep{2016ApJ...818L..22A}.  Therefore, understanding the nature of globular clusters that are likely to contain black holes is extremely important for investigating the conditions under which black hole binaries evolve.

\vspace{-1mm}

\section{Input for the High Latitude Wide Area Survey}

\vspace{-1mm}

Our main recommendation for the HLWA survey would be to request that the fields observed in the survey include a variety of diverse galaxy types (e.g. spiral, lenticular, elliptical, dwarf) so that the properties of their globular cluster systems can be compared.   The current depth of the proposed survey is in line with our science goals, as we do not require the Roman survey to go significantly deeper than the LSST Wide-Fast-Deep field ($i$ band limit of 26.3 mag). The currently planned survey would reach a depth of 25.8-26.7 AB mag. 

\vspace{-1mm}

\subsection{Field}

We advocate for a range of different galaxy types (elliptical, spiral, dwarf) within 100 Mpc also covered in the Rubin/LSST footprint to be observed as part of the Roman HLWA survey. We especially recommend the inclusion of galaxies with extensive globular cluster systems, e.g., massive galaxies in the Virgo and Fornax galaxy clusters. A roughly circular 1700 square degree field for the HLWA centered on the Fornax Cluster (which includes several elliptical and lenticular type galaxies, as well as 800 dwarf and low surface brightness galaxies) would not only cover the second closest (20 Mpc) galaxy cluster but would include a number of nearby ($< 30$ Mpc) galaxy groups and field galaxies. Such a range of galactic environments would include a range of galaxy masses and morphologies and 
is a high latitude field that lies within the Rubin Wide Fast Deep low dust footprint and contains two of the Rubin and Euclid deep drilling fields (the Chandra Deep Field-South and the Euclid Deep Field South).
In the foreground, this field contains four Milky Way globular clusters (AM-1, Eridanus, NGC 1261, and NGC 1851), the Fornax dwarf spheroidal galaxy (a Milky Way satellite that hosts six globular clusters), as well as seven fainter Milky Way satellites.

\vspace{-1mm}

\subsection{Filters}
\vspace{-1mm}


At 30 Mpc, a typical GC has an AB mag of 24.4 in $i$ and 24.0 in F158, so matching the LSST WFD depth with Roman would be ideal. For our science goals, the planned filters listed in order of most helpful to least helpful are: F158, F129, F184, F106.  However, we strongly advocate for the addition of the F213 filter, which is the most important filter for distinguishing globular clusters from Milky Way stars \citep[e.g.][]{2014ApJS..210....4M, 2022ApJ...930...24G}, and which is not currently included in the HWLA survey and goes further to the red than the reddest of the Euclid filters. 


\vspace{-1mm}

\section{Synergies with other science areas}
\vspace{-1mm}

Although we have specifically focused on the galaxies and globular cluster systems that will be covered by Rubin/LSST,  covering a large sample of galaxies within 100 Mpc would immediately broaden the range of science that can be done with the combined Roman + Rubin data.  For example, this would expand the galaxy sample to include low surface brightness galaxies, which have also been shown to host extensive globular cluster populations.  This would also enable measurements of accurate distances to many thousands of galaxies out to at least 100 Mpc.   Also relevant here is the white paper led by Montes et al., which discusses the science that can be done with diffuse intracluster light in galaxy clusters and groups, the white paper from Bechtel et al., which discusses globular cluster stellar streams.  We therefore advocate including additional star cluster systems in the Roman survey.  The resultant data set would constitute an invaluable resource for studies of galactic structure, near-field cosmology, and other areas of extragalactic science  \citep{2009ApJ...694..556B, 2018ApJ...854L..31C, 2021ApJ...911...65B}.

Moreover, Roman will be able to extend previous work on young star clusters in nearby galaxies carried out using the Hubble and James Webb Space Telescopes to a much larger sample of galaxies, for example by determining the cluster mass function and its variation with age \citep{mok2019} and the fraction of stars formed in clusters \citep{Cook2023}. The HLWA will allow studies on how these vary between and within galaxies.

\bibliography{clusters}
\bibliographystyle{aasjournal}

\end{document}